\documentclass[pre,aps,twocolumn,showpacs]{revtex4}
\usepackage{epsfig}
\usepackage{graphicx}
\usepackage{textcomp}

\begin{document}

\title{On population extinction risk in the aftermath of a catastrophic event}
\author{Michael Assaf$^1$, Alex Kamenev$^2$, Baruch Meerson$^1$}
\affiliation{$^1$ Racah Institute of Physics, Hebrew University of
Jerusalem, Jerusalem 91904, Israel\\$^2$ Department of Physics,
University of Minnesota, Minneapolis, Minnesota 55455, USA} 

\begin{abstract}
We investigate how a catastrophic event (modeled as a temporary fall of the reproduction rate)
increases the extinction
probability of an isolated self-regulated stochastic population. Using a variant of the
Verhulst logistic model as an example, we combine the probability
generating function technique with an eikonal approximation to
evaluate the exponentially large increase in the extinction
probability caused by the catastrophe. This quantity is given by the
eikonal action computed over ``the optimal path" (instanton) of an
effective classical Hamiltonian system with a time-dependent
Hamiltonian. For a general catastrophe the eikonal equations can be
solved numerically. For simple models of catastrophic events
analytic solutions can be obtained. One such solution becomes quite
simple close to the bifurcation point of the Verhulst model. The
eikonal results for the increase in the extinction probability
caused by a catastrophe agree well with numerical solutions of the
master equation.

\end{abstract}
\maketitle

\section{Introduction}
\label{intro}

\begin{figure}
\includegraphics[width=5.6cm,height=4.5cm,clip=]{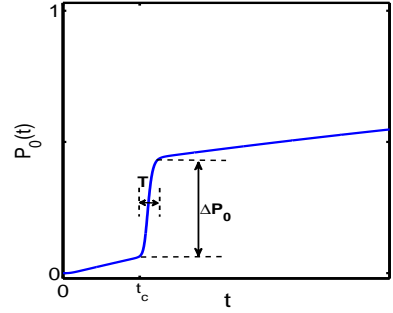}
\caption{(Color online) A schematic plot of the time-dependent
extinction probability ${\cal P}_0(t)$ showing the effect of a
catastrophe with a characteristic duration $T$.} \label{sketch}
\end{figure}

Evaluation of the extinction risk of a population in the aftermath
of a catastrophe - a drastic deterioration of environmental
conditions - is of great importance in population biology
\cite{book}. A practical approach to modeling of this problem
assumes that (i) the population dynamics is governed by a (Markov)
stochastic birth-death process \cite{Gardiner,vankampen}, (ii) this process has
an absorbing state at zero population size, and (iii) the
catastrophe introduces an explicit (and possibly strong) time
dependence into one or more of the transition rates. The question we
want to address in this work is the following: What is the
extinction probability of the population at the time the
catastrophic event is over and the environmental conditions return
to normal? The explicit time dependence of the transition rates,
brought about by the catastrophe, makes the problem difficult for
analysis. We will present here a formalism that helps develop an
insight into this class of problems. As a prototypical example of
the population dynamics we will adopt a variant of the stochastic
Verhulst logistic model. We will be interested in the regime of
parameters where, if there is no catastrophe, a long-lived
quasi-stationary distribution of the population sizes is formed.
Because of the presence of an absorbing state at zero the population
ultimately goes extinct in this model even without a catastrophe. A
natural measure of the extinction \textit{dynamics} is ${\cal P}_0(t)$: the
probability that the population goes extinct by the time $t$. The
focus of attention in this work is on the \textit{increase} in the
extinction probability ${\cal P}_0(t)$ caused by the catastrophe,
see Fig. \ref{sketch}. We will assume that, if there is no
catastrophe, the mean time to extinction (MTE), although finite, is too long to be
relevant. Let ${\cal P}_0^{B}(t)$ be the extinction probability
before the catastrophe occurred. Neglecting a transient related to
the relaxation  to the quasi-stationary distribution (and the
exponentially small extinction probability during this stage), this
probability grows very slowly according to ${\cal P}_0^{B}(t)\simeq
1-e^{-t/\tau}$, see \textit{e.g.} \cite{Kamenev1,Assaf2,Assaf3}.
Here $\tau$ is the MTE if there is no catastrophe. During the
catastrophe, which starts at $t=t_c$, the extinction probability
${\cal P}_0(t)$ grows more rapidly and, after the catastrophe,
reaches a value ${\cal P}_0^*$. From then on (again, neglecting a
relaxation transient), the extinction probability grows very slowly,
approximately as ${\cal P}_0^A(t)\simeq 1-(1-{\cal
P}_0^*)e^{(t_c-t)/\tau}$. We assume that the effect of the
catastrophe is not too weak, ${\cal P}_0^* \gg {\cal P}_0^{B}$,
where ${\cal P}_0^{B}$ is the (exponentially small) accumulated
extinction probability before the catastrophe.  The goal of this
work is to evaluate the increase of the extinction probability
$\Delta{\cal P}_0\equiv {\cal P}_0^*-{\cal P}_0^{B}$.

The formalism that we  will employ to achieve this goal begins from
a transformation of the master equation  (with time-dependent
transition rates)  into an exact evolution equation for the
probability generating function. For the problem in question this
equation is a second-order linear partial differential equation
(PDE). We then apply a time-dependent eikonal approximation to this
evolution equation. This approximation, in many ways similar to the
time-dependent WKB approximation in quantum theory \cite{LL}, brings
the second-order PDE to a first-order PDE of Hamilton-Jacobi type,
with an effective classical Hamiltonian which explicitly depends on
time. The further analysis deals with the characteristic lines of
the Hamilton-Jacobi equations: the phase trajectories generated by
the effective classical Hamiltonian.

We will assume throughout the paper that both the initial population
size, and the expected population size immediately after the
catastrophic event, are much larger than unity. Under this
assumption (and not too close to the bifurcation point of the
Verhulst model, see below) we will evaluate the (exponentially
large) net increase $\Delta{\cal P}_0$ in the extinction probability
caused by the catastrophe. This formalism yields an asymptotically
correct value of the corresponding large exponent which makes it
advantageous compared to the widely used van Kampen's system size
expansion of the master equation  and related methods, see
\textit{e.g.} \cite{Gardiner,vankampen}. Indeed, as it is well
established by now, the Fokker-Planck equation, resulting from the
van Kampen's system size expansion, cannot predict the MTE of
stochastic birth-death systems with exponential accuracy. This can
be traced to the failure of the Fokker-Planck equation in describing
the actual probability distribution tails of these systems
\cite{Gaveau,Kamenev1,Doering,Assaf2,Assaf3}.

Here is the layout of the remainder of the paper. We begin Sec.
\ref{model} with a brief reminder on the stochastic Verhulst
logistic model. Then we derive the evolution equation for the
probability generating function and employ an eikonal approximation
in order to determine the MTE of the population if there is no
catastrophe. Here our results coincide, with exponential accuracy,
with those of previous works on the stochastic Verhulst logistic
model. Section \ref{catastrophe} deals with the effect of a
catastrophic event on the population survival. We first demonstrate
the efficiency of the eikonal method by finding numerically the most
probable path to extinction, and computing the corresponding
increase in the extinction probability due to the catastrophe, for a
typical example of a catastrophic event. Then we obtain
non-perturbative (in the catastrophe magnitude) analytical results
by adopting a simple schematic form of the catastrophic event: we
postulate that the reproduction rate of the population drops
instantaneously to zero at a specified time and recovers to the
pre-catastrophe value, again instantaneously,  after a given time
$T$ has elapsed. Section \ref{numerics} compares our numerical and
analytical eikonal results for the extinction probability with
direct numerical solutions of the master equation. A brief summary
and discussion of our results is presented in Sec. \ref{discussion}.

\section{Verhulst model, probability generating function and eikonal approximation}
\label{model} As a prototypical example of self-regulating  dynamics
of an isolated population we consider a variant of the
Verhulst logistic model: a Markov single-step stochastic birth-death
process with continuous time. If there is no catastrophe,  the reproduction and mortality
rates are given by
\begin{equation}\label{verhulst}
\lambda_n= B\,n\;\;\;\mbox{and}\;\;\;\mu_n=n+\frac{Bn^2}{N}\,,
\end{equation}
respectively, where we use the same notations as in Ref.
\cite{Doering}. The reproduction rate per individual is constant,
while the mortality rate per individual is constant at small population
sizes, but grows proportionally to the population size when the
latter is sufficiently large, accounting, for example, for
competition for resources. For brevity, time and the transition rates in
Eq.~(\ref{verhulst}) are rescaled with respect to the value of the
mortality rate at small population sizes.

At the level of deterministic modeling, the dynamics of the
population size is described by the rate equation
\begin{equation}\label{rateeq}
\dot{n}(t)= (B -1)\,n(t)-\frac{B}{N}\, n^2(t) \,.
\end{equation}
For $B>1$  this equation has an attracting fixed point
$n_s=N(1-1/B)$  and a repelling fixed point $n_0=0$.  Throughout the
paper we assume $n_s \gg 1$; this necessarily requires $N\gg 1$.
A linear stability analysis of Eq.~(\ref{rateeq}) around $n=n_s$ yields the
characteristic relaxation time $\tau_0=(B-1)^{-1}$ (in the units of
the mortality rate at small population sizes).

Demographic stochasticity in the Verhulst model is accounted for by the master equation
\begin{eqnarray}
&&\frac{d{\cal P}_n}{d t} = B (n-1){\cal P}_{n-1} - Bn {\cal P}_n \nonumber\\
&+& \left[ n+1+\frac{B (n+1)^2}{N}\right] {\cal P}_{n+1}
-\left(n+\frac{Bn^2}{N}\right){\cal P}_n  ,\label{master}
\end{eqnarray}
where ${\cal P}_n(t)$ is the probability that the population size at
time $t$ is $n$. The stochastic Verhulst model, as described by
Eq.~(\ref{master}), and related models were considered in many
works, see \cite{Doering,weiss,Barbour,oppenheim,kryscio,Nasell01}
and references therein. It was found that, under certain conditions
specified below, a long-lived quasi-stationary distribution,
conditioned on non-extinction, is formed  in this system after the
relaxation time ${\cal O}(\tau_0)$. The quasi-stationary
distribution has a peak with relative width proportional to
$N^{-1/2}$ around the attracting state $n_s$ of the deterministic
description. At much longer times, however, the population goes
extinct. This is because $n=0$ is an absorbing state, so a rare
sequence of events brings the process there with certainty. We will
work in such a parameter regime that, if there is no catastrophe,
the MTE is too long to be relevant. That is, we will be interested
in times which, although much longer than the relaxation time
$\tau_0$, are still much shorter than the MTE of the system without
a catastrophe, that is the one described by Eq.~(\ref{master}).

Notice that the same master equation (\ref{master}) also describes
the stochastic dynamics of three
chemical reactions: $A\stackrel{\lambda}{\rightarrow} 2A$,\,
$A\stackrel{\mu}{\rightarrow} \emptyset$, and
$2A\stackrel{\sigma}{\rightarrow} A$ \cite{Kamenev2}, with rates
$\lambda \equiv B\,$, $\mu \equiv 1+B/N$, and $\sigma=2(\mu-1)$,
respectively.  For definiteness we will use the notation of the Verhulst model
in the following.

We will first demonstrate the method on the (well known) case of population extinction
when there is no
catastrophe. Introduce the probability generating function
\cite{Gardiner,vankampen}
\begin{equation}
G(\wp,t)=\sum_{n=0}^{\infty} \wp^n {\cal P}_n(t)\,,
\end{equation}
where $\wp$ is an auxiliary variable. $G(\wp,t)$ encodes all the
probabilities ${\cal P}_n(t)$, as those are given by the coefficients
of its Taylor expansion around $\wp=0$. The generating function obeys the
normalization condition
\begin{equation}\label{norm}
    G(1,t)=1
\end{equation}
which follows from the conservation of the total probability. The distribution
moments can be expressed through the $\wp$-derivatives
of the generating function at $\wp=1$, \textit{e.g.} $\langle
n\rangle(t) \equiv \sum_n n {\cal P}_n(t) = \left.\partial_\wp
G(\wp,t)\right|_{\wp=1}$.

By multiplying Eq.~(\ref{master}) by $\wp^n$ and summing over all
$n$ one obtains, after some algebra, an evolution equation for
$G(\wp,t)$:
\begin{equation}\label{Gdot1}
\frac{\partial G}{\partial t} =
\frac{B}{N}\left(1-\wp\right)\wp\,\frac{\partial^2 G}{\partial
\wp^2}+ (\wp-1)\left(B\wp-1-\frac{B}{N}\right)\frac{\partial
G}{\partial \wp}\,.
\end{equation}
The signature of the (empty) absorbing state is the absence of
a term proportional to $G(\wp,t)$ in Eq.~(\ref{Gdot1}). This immediately
brings about the stationary solution $G(\wp,t)=1$ corresponding to the empty state.

In contrast to the Fokker-Planck equation, which is derivable from
the master equation (\ref{master}) by the van Kampen's system size
expansion, the starting point of our theory - the evolution equation
(\ref{Gdot1}) - is exact. Throughout the rest of the paper we assume $N\gg B$. As our
main results below are obtained with exponential accuracy, this strong
inequality enables us to neglect the term
$(B/N)\partial G/\partial \wp$,
and Eq.~(\ref{Gdot1}) becomes
\begin{equation}\label{Gdot}
\frac{\partial G}{\partial t} =
\frac{B}{N}\left(1-\wp\right)\wp\,\frac{\partial^2 G}{\partial
\wp^2}+ (\wp-1)\left(B\wp-1\right)\frac{\partial G}{\partial \wp}\,.
\end{equation}
This second-order PDE can be interpreted as an imaginary-time
Schr\"{o}dinger equation $\partial G/\partial t=\hat{{\cal H}}G$,
where
$$
\hat{{\cal H}} =
\frac{B}{N}\left(1-\wp\right)\wp\,\frac{\partial^2}{\partial \wp^2}+
(\wp-1)\left(B\wp-1\right)\frac{\partial}{\partial \wp}
$$
is the Hamiltonian operator.

In the framework of the spectral theory \cite{Assaf1,Assaf2,Assaf3},
the solution of the initial value problem for Eq.~(\ref{Gdot}) can
be obtained via an expansion in the eigenfunctions of a
Sturm-Liouville problem related to the non-Hermitian operator
$\hat{{\cal H}}$. The boundary conditions, needed for this
Sturm-Liouville problem, come from the demand of boundedness of the
eigenfunctions at the singular points of $\hat{{\cal H}}$. In this
way one obtains a zero eigenvalue $E_0=0$ which describes the true
(absorbing) steady state of the system: extinction of the population. In addition, one
obtains a discrete set of negative eigenvalues
$\{E_n\}_{n=1}^{\infty}$ that describe the decay with time of the
discrete set of eigenfunctions. The relative weight of each
eigenfunction in the expansion is determined,  via $G(\wp,t=0)$, by
the initial probability distribution ${\cal P}_n(t=0)$. For $N \gg
1$ (and not too close to the bifurcation point $B=1$) the higher
eigenvalues $E_2,E_3,\dots$ scale as $\tau_0^{-1}$, whereas the
fundamental eigenvalue $E_1$ is exponentially small. Therefore,
there are two widely separated time scales in the problem. During
the fast relaxation time ${\cal O}(\tau_0)$ the higher modes decay
exponentially, and the probability distribution approaches the
quasi-stationary distribution (QSD) mentioned above, which is
derivable from the fundamental eigenfunction. The exponentially
small decay rate of the fundamental mode (and, correspondingly, of
the time-dependent metastable distribution that is proportional to
the QSD) is equal to the inverse MTE of the system. We checked that,
as in other similar systems \citep{Assaf2,Assaf3,Assaf1}, the
spectral theory yields accurate results both for the complete
probability distribution and for the MTE, including the
pre-exponent, in the catastrophe-free regime. However, the spectral
theory cannot, in general, accommodate time-dependent transition rates: the main
subject of this work. Therefore, in what follows we will use instead
the time-dependent eikonal approximation \citep{Kamenev1}. Although
it only gives exponential accuracy, the eikonal approximation is
readily generalizable to multi-population problems \citep{KM} and, as
we will demonstrate shortly, to time-dependent Hamiltonians.

Now we use the eikonal ansatz $G(\wp,t)=\exp[-{\cal S}(\wp,t)]$ in
the evolution equation (\ref{Gdot}). As can be checked \textit{a
posteriori}, ${\cal S}$ can be written as ${\cal S}=Ns(p,t)$, where
$s(p,t)={\cal O}(1)$. Therefore, the term
$(B/N)(1-\wp)\wp\,\partial_{pp}{\cal S}={\cal O}(1)$ is small compared to
the other terms which scale as $N$.  Neglecting it, we arrive at a first-order PDE
for $S(\wp,t)$
\begin{equation} \label{HJeq}
\frac{\partial {\cal S}}{\partial t}+\frac{B}{N}
\left(1-\wp\right)\wp\left(\frac{\partial {\cal S}}{\partial
\wp}\right)^2-(\wp-1)\left(B\wp-1\right)\frac{\partial {\cal
S}}{\partial \wp}=0\,,
\end{equation}
that can be interpreted as a Hamilton-Jacobi equation in the
momentum representation, where $\wp$ is the momentum.  Introducing the canonically conjugate coordinate $q=-\partial {\cal
S}/\partial \wp$, we obtain a one-dimensional classical
Hamiltonian flow with the time-independent Hamiltonian
\begin{equation}\label{H}
{\cal H}(\wp,q)=(1-\wp) \left(\frac{B}{N}\wp q -B\wp+1\right) q\,.
\end{equation}
The fact that ${\cal H}(1,q)=0$ reflects the probability
conservation, see Eq.~(\ref{norm}), whereas the fact that ${\cal
H}(\wp,0)=0$ manifests the presence of the absorbing empty
state.

\begin{figure}
\includegraphics*[width=6.5cm,height=5.6cm,clip=]{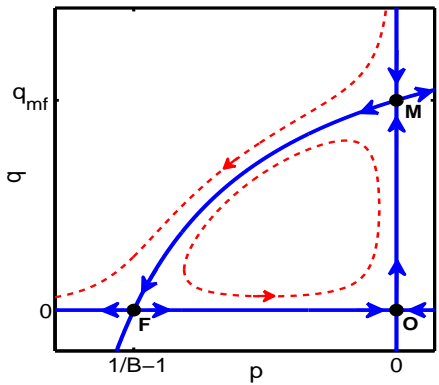}
\caption{(Color online) The phase plane $(p,q)$ emerging in the
eikonal approximation to the stochastic Verhulst model. The fixed
points are denoted by $M$ (the mean-field point), $F$ (the
fluctuational point) and $0$ (the trivial point). Solid lines depict
zero-energy trajectories, including the heteroclinic trajectory
$MF$. Dashed lines show two examples of non-zero-energy
trajectories.} \label{phaseplane1}
\end{figure}

It is convenient to shift the momentum  $p=\wp-1$, leaving the
coordinate $q$ unchanged. The new Hamiltonian is
\begin{equation}\label{H1}
H(p,q)=p\left[-\frac{B}{N}(p+1)q +B (p  + 1)-1 \right]\, q\,,
\end{equation}
while the Hamilton equations are
\begin{eqnarray}
  \dot{q} &=& \frac{\partial H}{\partial p} =  (2p+1)\left(B q -\frac{B}{N} q^2 \right)  -q  \,, \label{qdot} \\
  \dot{p} &=& -\frac{\partial H}{\partial q} = -(p^2+p) \left(B - \frac{2B}{N}  q  \right)+ p \label{pdot}\,.
\end{eqnarray}
This Hamiltonian flow was investigated, in the context of the three
above-mentioned chemical reactions, by Elgart and Kamenev
\cite{Kamenev2}. The phase plane $(p,q)$, defined by the Hamiltonian
(\ref{H1}), provides a useful visualization of the extinction
dynamics, see Fig. \ref{phaseplane1}. As the Hamiltonian does not
depend explicitly on time, it is an integral of motion:
$H(p,q)=E=const$, and the problem is integrable.  The \textit{zero
energy} trajectories, $E=0$, play a special role here, as will
become clear shortly. One type of zero-energy trajectories are the
mean field trajectories, staying on the line $p=0$. Here
Eq.~(\ref{qdot}) becomes
\begin{equation}\label{mfeq1}
\dot{q}= (B -1)\,q -\frac{B}{N}\, q^2
\end{equation}
which coincides with the rate equation (\ref{rateeq}). This fact
provides the interpretation of the $q$-variable as the reaction
coordinate.  Importantly, the attracting fixed point of the
mean-field,  or deterministic line $p=0$, $q_{mf}\equiv
n_s=N(1-1/B)$ becomes a hyperbolic point $(0,q_{mf})$ of the phase
plane $(p,q)$. We call this fixed point the mean-field point and
denote it by $M$.

The Hamiltonian (\ref{H1}) has two more zero-energy lines. One of
them is the extinction line $q=0$ which includes two more hyperbolic
fixed points of the phase plane $(p,q)$: the ``fluctuational point"
$(1/B-1,0)$, denoted by $F$, and the trivial point $(0,0)$ denoted
by $0$.   The third zero-energy curve
\begin{equation}\label{q(p)}
q= q_0(p)=N-\frac{N}{B(p+1)}\,,
\end{equation}
includes a heteroclinic trajectory that exits, at time $t=-\infty$,
the mean field point $M$ along its unstable eigendirection, and
enters, at time $t=\infty$, the fluctuational point $F$ along its
stable eigendirection. This heteroclinic trajectory represents the
optimal (most probable) path, or the instanton connection, of the
extinction dynamics. Indeed, it describes, in the eikonal language,
the most probable sequence of discrete events bringing the system
from its quasi-stationary state to extinction, \textit{cf.}
\cite{freidlin,Graham,Dykman1,Kamenev1}. If there is no catastrophe,
the MTE of the population is, with exponential accuracy,  $\tau \sim
\exp({\cal S}_0)$, where the (zero-energy) action is
\begin{equation}\label{action1}
    {\cal S}_0 = -\int_{-\infty}^{\infty} q\,\dot{p}\,dt \,,
\end{equation}
and the integration should be performed along the zero-energy
heteroclinic trajectory (\ref{q(p)}). The result is
\begin{equation}\label{SV}
\tau \sim \exp\left[-\int_0^{\frac{1}{B}-1}
q_0(p) \,dp\right]=\exp\left(N\,\frac{B-1-\ln B}{B}\right)\,.
\end{equation}
The approximation is valid when ${\cal S}_0 \gg 1$. One can check that the
result (\ref{SV}) coincides, up to a pre-exponent, with those of
\cite{Doering,Barbour,Nasell01,spectral}.

Of a special interest is the region of parameters close to the bifurcation point of the Verhulst model:
$N^{-1/2} \ll B-1 \ll 1$, where the left inequality is required for
the validity of the eikonal approximation.  For $B-1\ll 1$ the
action
\begin{equation}\label{area}
{\cal S}_0 \simeq \frac{N (B-1)^2}{2}\,,
\end{equation}
scales as the square of the distance to the bifurcation point. It
has been found recently \cite{Dykman2,KM} that, close to the
bifurcation point, the Verhulst model, the SIS (Susceptible -
Infected - Susceptible) model of epidemiology
\cite{weiss,Barbour,oppenheim,kryscio,Nasell01,Doering}, the SIS
model with demography \cite{Nasell02,Dykman2}, the SIR (Susceptible
- Infected -Recovered) model with demography
\cite{Herwaarden,Nasell99,KM} and other related stochastic models
become, in the leading order, \textit{identical} if their rates are
properly rescaled. One can check that, in this limit, the term
$-(B/N)pq $ in the square brackets in Eq.~(\ref{H1}) can be
neglected compared to the rest of the terms, and one arrives at the
``universal" Hamiltonian
\begin{equation}\label{H1univ}
H(p,q)=p\left(-\frac{q}{N} +p+ B-1\right)\,q\,,
\end{equation}
introduced in Ref. \cite{Kamenev2} in the context of the three chemical reactions mentioned above.  All
three zero-energy lines in the phase plane of the universal
Hamiltonian are straight lines, and the action ${\cal S}_0$, given by
Eq.~(\ref{area}), is simply the area of
the triangle formed by these straight lines.

\section{Catastrophic event and action calculation}\label{catastrophe}
To model a catastrophic event we assume that, because of unfavorable
environmental changes, the reproduction rate drops and then recovers
to the pre-catastrophe value. This can be described by introducing a time-dependent factor $f(t)$, such that $f(\pm\infty)=1$, into
the reproduction rate:
\begin{equation}\label{catastrophe1}
\lambda_n(t)= B f(t)\,n\,,
\end{equation}
The mortality rate $\mu_n$ remains constant in time in this model, see Eq.~(\ref{verhulst}). At the
level of deterministic modeling, the population size is
described by the rate equation
\begin{equation}\label{rateeq1}
\dot{n}(t)= \left[B f(t)-1\right]\,n(t)  -\frac{B}{N}\,n^2(t) \,,
\end{equation}
the master equation becomes
\begin{eqnarray}
&&\hspace{-5mm}\frac{d{\cal P}_n}{d t} = B f(t) (n-1){\cal P}_{n-1} - B f(t) n {\cal P}_n \nonumber\\
&&\hspace{-5mm}+   \left[ n+1+\frac{B (n+1)^2}{N}\right]{\cal
P}_{n+1} -\left(n+\frac{Bn^2}{N}\right)  {\cal P}_{n}
,\label{master1}
\end{eqnarray}
while the evolution equation for $G(\wp,t)$ is again an
imaginary-time Schr\"{o}dinger equation $\partial G/\partial
t=\hat{{\cal H}}G$ with the Hamiltonian operator
$$
\hat{{\cal H}} =
\frac{B}{N}\left(1-\wp\right)\wp\,\frac{\partial^2}{\partial \wp^2}+
(\wp-1)\left[B f(t) \wp-1\right]\frac{\partial}{\partial \wp}
$$
that now explicitly depends on time. The same eikonal ansatz
$G(\wp,t)=\exp[-{\cal S}(\wp,t)]$ brings about the Hamilton-Jacobi equation
for ${\cal S}(\wp,t)$ which defines a
classical Hamiltonian
flow with the time-dependent Hamiltonian
\begin{equation}\label{H1time}
H(p,q,t)=p\left[-\frac{B}{N}(p+1)q +B f(t) (p  + 1)-1 \right] q\,,
\end{equation}
with the same coordinate $q$ and shifted momentum  $p=\wp-1$ as in the
time-independent case. The Hamilton equations are
\begin{eqnarray}
  \dot{q} &=& \frac{\partial H}{\partial p} = (2 p  +1)\left[B f(t)q-\frac{B}{N} q^2 \right]  - q  \,, \label{qdot1} \\
  \dot{p} &=& -\frac{\partial H}{\partial q} = - (p^2+p) \left[B f(t) - \frac{2B}{N}  q  \right]+ p \label{pdot1}\,.
\end{eqnarray}
As the Hamiltonian $H$ is not an integral of motion anymore, the
problem is in general non-integrable.  There are, however, two
planes in the extended phase space $(p,q,t)$ where the Hamiltonian
is still conserved and equal to zero. These are the mean-field
plane $(p=0,q,t)$ and the extinction plane $(p,q=0,t)$.  In the
mean-field plane Eq.~(\ref{qdot}) becomes
\begin{equation}\label{mfeqtime}
\dot{q}=\left[B f(t) -1\right] q -\frac{B}{N} q^2
\end{equation}
which again coincides with the rate equation (\ref{rateeq1}). In the
eikonal approximation the extinction occurs along the instanton
connection of the time-dependent Hamiltonian (\ref{H1time}) in the
extended phase space $(p,q,t)$. To reasonably describe a catastrophe
the function $f(t)$ should be bounded from above. In this case there
exists a heteroclinic trajectory $[p_{op}(t),q_{op}(t)]$ which exits
the mean-field fixed point $M$ well before the catastrophe and
arrives at the fluctuational fixed point $F$ after the catastrophe
is over. The full action along this heteroclinic trajectory
\begin{equation}
{\cal
S}=\int_{-\infty}^{\infty}\left\{-q_{op}(t)\dot{p}_{op}(t)-H[q_{op}(t),p_{op}(t),t]\right\}dt\,
\label{totacttheo}
\end{equation}
[where $H$ is given by Eq.~(\ref{H1time})], determines (the
logarithm of) the increase in the extinction probability caused by
the catastrophe.

\subsection{Instanton calculations: numerical solution}
For a smooth $f(t)$, the Hamilton equations (\ref{qdot1}) and
(\ref{pdot1}) (and similar Hamilton equations for other models) can
be accurately solved by a shooting method with a single shooting
parameter. As in the unperturbed case, the instanton connection must
exit, at $t=-\infty$, the  mean-field point $M$ and enter, at
$t=\infty$, the fluctuational point $F$. In a numerical solution we
specify, at some initial moment of time $t=t_{in}$ prior to the
catastrophe, the coordinate $q(t_{in})$ and momentum $p(t_{in})<0$
which lie on the \textit{unperturbed} heteroclinic trajectory
(\ref{q(p)}) in a vicinity of the mean-field point $M$. The initial
momentum $p(t_{in})$ can play the role of the shooting parameter. It
must be chosen sufficiently close to zero for the numerically found
instanton to approximate well the entire perturbed instanton.
On the other hand, the instanton must reach, at a final time $t_f$
after the catastrophe, (a close vicinity of) the fluctuational point
$F$. Therefore, for too a small $p(t_{in})$, intrinsic logarithmic
slowdown near the fixed point $t_f$ increases the
computation time and causes accumulation of numerical errors.
Denoting the numerically found instanton (or optimal
path) by $[p_{op}^{(n)}(t),q_{op}^{(n)}(t)]$, we can write the
action along it as
\begin{equation}
{\cal
S}_{num}=\int_{t_{in}}^{t_f}\left\{-q_{op}^{(n)}(t)\dot{p}_{op}^{(n)}(t)-H\left[q_{op}^{(n)}(t),p_{op}^{(n)}(t),t\right]\right\}dt.
\label{totact}
\end{equation}

\begin{figure}
\includegraphics[width=6.5cm,height=5.6cm,clip=]{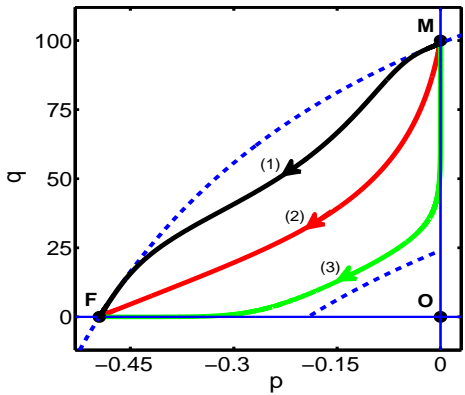}
\caption{(Color online) Projections on the $(p,q)$ plane of the ``extinction instantons" found by solving
the Hamilton equations (\ref{qdot1}) and (\ref{pdot1}) numerically
for $f(t)$ given by Eq.~(\ref{gencat}). The parameters are $N=200$,
$B=2$, $\Delta B=0.75$, and $t_c-t_{in}=40$. The catastrophe
durations are $T=1$ (solid line 1), $T=3$ (solid line 2) and $T=20$
(solid line 3). The upper dashed line is the instanton when there is
no catastrophe and $f(t)=1=const$. The lower dashed line is the
instanton for the system with $B$ replaced by $B-\Delta B = const$.
The upper and lower dashed lines yield the upper and lower bounds,
respectively, on the action as a function of $T$.} \label{nums}
\end{figure}

As an example, let us consider
\begin{equation}\label{gencat}
f(t)=1-\frac{\Delta B}{B} \, e^{-\frac{(t-t_c)^2}{T^2}}\,,
\end{equation}
where $\Delta B \leq B$ is the catastrophe magnitude as manifested
in the change of the reproduction rate, $t_c$ is the time when the
catastrophe reaches its maximum, and $T$ is the catastrophe
duration. Figure \ref{nums} shows projections on the $(p,q)$ plane of the
numerically found instantons
for several values of the catastrophe duration $T$. We compared the
numerically found action (\ref{totact}) to (the logarithm of) the
increase of the extinction probability due to the catastrophe,
$\Delta {\cal P}_0$, determined by solving numerically the master
equation (\ref{master1}) with $f(t)$ from Eq.~(\ref{gencat}). Such a
comparison is presented in Sec. $4$.

\subsection{Instanton calculations: analytically soluble example}
\begin{figure}
\includegraphics[width=5.5cm,height=4.5cm,clip=]{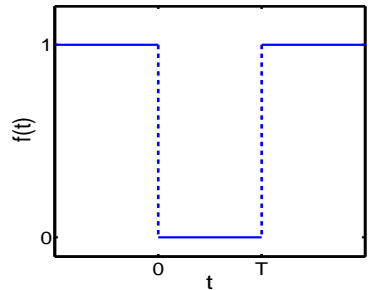}
\caption{(Color online) A model catastrophic event for which the
eikonal problem is exactly soluble. The catastrophe starts at time
$t=0$, when the reproduction rate drops to zero, and ends at $t=T$,
when the reproduction rate recovers to its pre-catastrophe value.}
\label{figlam}
\end{figure}

\begin{figure}
\includegraphics[width=5.3cm,height=4.3cm,clip=]{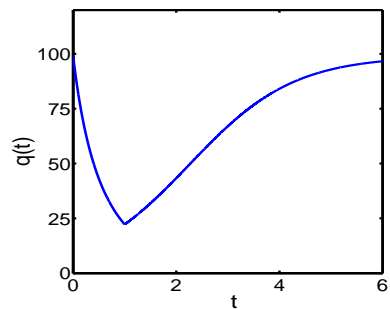}
\caption{(Color online) Effect of the catastrophic event, described
by Eq.~(\ref{f}), on the expected population size as predicted by
the solution (\ref{mfcatast}) of the deterministic rate equation
(\ref{mfeqtime}). The parameters are $N=200$, $B=2$, and $T=1$.}
\label{detdynamics}
\end{figure}

We  present now an example when the eikonal equations, describing the impact of a catastrophe on the population extinction probability, are exactly soluble. We assume that the reproduction
rate $\lambda_n$ drops instantaneously to zero at $t=0$ and
recovers to the pre-catastrophe value, again instantaneously,  after
time $T$ has elapsed. In terms of the function $f(t)$ we have
\begin{equation}\label{f}
f(t)=\left\{ \begin{array}{ll}
1 & \;\;\;\mbox{if $\;t<0$ or $t>T$}\,, \\
0 & \;\;\;\mbox{if $\;0<t<T$\,,}
\end{array}
\right.
\end{equation}
see Fig. \ref{figlam}.
With this $f(t)$ the solution of the deterministic rate equation
(\ref{mfeqtime}) is
\begin{equation}\label{mfcatast}
n(t)= n_s \times \left\{ \begin{array}{ll}
\left[B (e^{ t}- 1)+1  \right]^{-1}\,,\hspace{1.9cm}0<t<T\,,\\\\
\left[B (e^{T}-1)e^{-(B-1)(t-T)}+1 \right]^{-1}\,, \;\;\;t>T\,,
\end{array}
\right.
\end{equation}
where $n(t=0)=n_s=N(1-1/B)$ corresponds to the mean-field point $M$.
The deterministic solution, shown in Fig. \ref{detdynamics}, predicts a complete
recovery of the population after the catastrophe. However, the stochastic effects, missed by the
rate equation, can be greatly enhanced because of the temporary decline in
the population size. As a result, they can
increase the extinction probability considerably. As we rely on the
eikonal approximation in describing this effect, we are interested
in the regime where, in spite of the catastrophe, the expected population
size by the end of the catastrophe, $n(T)$, is still sufficiently large: $n(T)
\gg 1$.

\begin{figure}
\includegraphics[width=6.5cm,height=5.6cm,clip=]{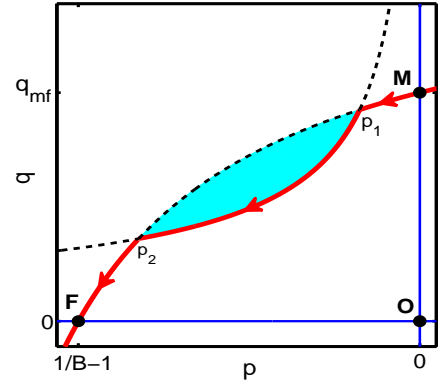}
\caption{(Color online) Projection on the $(p,q)$ plane of the
``extinction instanton" (the thick solid line going from $M$ to $F$)
for the catastrophic event described by Eq.~(\ref{f}). Points $p_1$
and $p_2$ correspond to times $t=0$ and $t=T$ where the catastrophic
event begins and ends, respectively. At $t<0$ and $t>T$ the
instanton follows the zero-energy heteroclinic trajectory
$q=q_0(p)$, whereas at $0<t<T$ it follows a non-zero-energy
trajectory $q=q_c(p,E_c)$. The energy $E_c=E_c(T)$ is determined by
the given catastrophe duration $T$. The area of the shaded region
corresponds to the integral term in Eq.~(\ref{deltaS}).}
\label{phasespace}
\end{figure}

Because of the special shape of the function $f(t)$, there are now two distinct
Hamiltonians: the unperturbed
Hamiltonian (\ref{H1}) before and after the catastrophe
and the zero-reproduction-rate Hamiltonian during the catastrophe:
\begin{equation}
H_c(p,q)=-p\left[\frac{B}{N}(p+1)q + 1 \right] q\,. \label{Hc}
\end{equation}
Each of the two Hamiltonians is an integral of motion on the
corresponding time interval. The instanton can be found by matching
three separate trajectory segments: the pre-catastrophe, catastrophe
and post-catastrophe segments. Figure \ref{phasespace} shows a
projection of the instanton on the $(p,q)$ plane. To recall, the
instanton must exit, at $t=-\infty$, the  mean-field point $M$ and
enter, at $t=\infty$, the fluctuational point $F$. The matching
conditions at times $t=0$ and $t=T$ are provided by the continuity
of the functions $q(t)$ and $p(t)$. The pre- and post-catastrophe
segments must have a zero energy, $E=0$, so they are parts of the
original zero-energy trajectory of the unperturbed problem, see
Eq.~(\ref{q(p)}).  For the catastrophe segment, however, the energy
$E=E_c$ is non-zero and \textit{a priori} unknown. It parametrizes
the intersection points $p_1$ and $p_2$ (see Fig. \ref{phasespace})
between the unperturbed zero-energy line
\begin{equation}
q_0(p)=N-\frac{N}{B(1+p)}\,,
\end{equation}
and the non-zero-energy line $H_c=E_c$:
\begin{eqnarray}
q_c(p,E_c)=\frac{N}{2B(1+p)}\left[\sqrt{1-\frac{4E_c(1+p)B}{N
p}}-1\right]. \label{qc}
\end{eqnarray}
Solving the algebraic equation $q_0(p)=q_c(p,E_c)$  for $p$, we
obtain
\begin{eqnarray}
p_1(E_c)&=&-\frac{B-1}{2B}\left[1- \sqrt{1-\frac{4E_c
B}{N(B-1)^2}}\,\right]\,,
\nonumber\\
p_2(E_c)&=&-\frac{B-1}{2B}\left[1+ \sqrt{1-\frac{4E_c
B}{N(B-1)^2}}\,\right]\,.
\end{eqnarray}
To determine $E_c$ we demand  the duration of the catastrophe to
be $T$.  Using Eq.~(\ref{pdot1}) for $f(t)=0$ and Eq.~(\ref{qc}),
we obtain an algebraic equation for $E_c=E_c(T)$:
\begin{equation}\label{T}
\int_{p_1(E_c)}^{p_2(E_c)}\frac{dp}{p\left[\,(2B/N) (p+1)\,
q_c(p,E_c) +1\,\right]}=T\,,
\end{equation}
where $q_c(p,E_c)$ is given by Eq.~(\ref{qc}). The net increase  of the extinction probability of the population due to the catastrophe
is proportional to $\exp[-{\cal S}(T)]$, where the action ${\cal
S}(T)$ is
\begin{eqnarray}
\label{S(T)}
 \nonumber {\cal S}(T) &=& \int_{1/B-1}^{p_2[E_c(T)]} q_0(p) \,dp + \int_{p_2[E_c(T)]}^{p_1[E_c(T)]} q_c[p,E_c(T)] \,dp \\
  &+& \int_{p_1[E_c(T)]}^0 q_0(p) \,dp  - E_c (T)\, T\,.
\end{eqnarray}
We can also rewrite this action as
\begin{eqnarray}
\label{deltaS} {\cal S}(T) &=& {\cal S}_0 -
\int_{p_2[E_c(T)]}^{p_1[E_c(T)]} \{q_0(p)-q_c[p,E_c(T)]\}\,dp
\nonumber\\ &-& E_c (T)\, T,
\end{eqnarray}
where ${\cal S}_0$ is the unperturbed action, see Eqs.~(\ref{action1}) and (\ref{SV}), and the
integral term in this equation corresponds to the area of the shaded
region in Fig. \ref{phasespace}.

To obtain more visual results, let us consider the limit of
$B-1\ll 1$. In this limit the pre- and post-catastrophe Hamiltonian reduces
to the universal Hamiltonian (\ref{H1univ}). Furthermore, as the
terms $(B/N)pq\,$ and $(B/N)q$ can be neglected here compared to $1$ in
Eq.~(\ref{Hc}), the zero-reproduction-rate Hamiltonian during the
catastrophe simplifies drastically:
\begin{equation}
H_c(p,q)\simeq -pq\,, \label{Hc1}
\end{equation}
and the catastrophe segment of the instanton trajectory becomes
simply
\begin{equation}\label{qcapprox}
q_c(p,E_c)\simeq-\frac{E_c}{ p}\,.
\end{equation}
The corresponding Hamilton equation for $p$ [Eq.~\ref{pdot1})], on
the catastrophe segment, is  $\dot{p}\simeq  p$, so $\ln
(p_2/p_1)\simeq  T$ and
\begin{equation}\label{ratio}
    \frac{p_2}{p_1}\simeq\exp( T)\,.
\end{equation}
In their turn, $p_1=p_1(E_c)$ and $p_2=p_2(E_c)$ are the roots of
the equation $q_0(p)=q_c(p,E_c)$ which now reads
\begin{equation}\label{quadratic}
N(B-1+p) \simeq - \frac{E_c}{p} \,.
\end{equation}
Equations~(\ref{ratio}) and (\ref{quadratic})  yield
\begin{equation}
p_1\simeq\frac{1-B}{e^{T}+1}\,,\;\;\;\;\;p_2\simeq\frac{(1-B)e^{T}}{e^{T}+1}\,,
\end{equation}
and
\begin{equation}
E_c \simeq \frac{N(B-1)^2}{4} \cosh^{-2} \left( \frac{T}{2}
\right)\,.
\end{equation}
One can see that, as $T\to 0$, the intersection points $p_1$ and $p_2$
merge signaling no change of the unperturbed action. As $T \to
\infty$, $p_1$ approaches the mean-field point $M$, whereas $p_2$
approaches the fluctuational point $F$. In the latter case the
population quickly dies out.

Now we use Eq.~(\ref{deltaS}) to calculate the
extinction action caused by the catastrophe.
After a simple algebra we obtain
\begin{equation}
\label{deltaSsimple}  {\cal S}(T) \simeq \frac{2{\cal
S}_0}{1+e^{T}}\,,
\end{equation}
where the unperturbed action ${\cal S}_0$ is given by Eq.~(\ref{area}). For short
catastrophe durations, $T\ll 1$, we obtain a small correction,
linear in $T$,  to the unperturbed action:
\begin{equation}
\label{deltaSshort} {\cal S}(T) \simeq {\cal S}_0(1-T/2) \, .
\end{equation}
For long catastrophes, $T\gg 1$, the total action ${\cal S}(T)$
decays exponentially with an increase of $T$,
\begin{equation}
\label{deltaSlong} {\cal S}(T) \simeq 2{\cal S}_0 \,e^{-T}\,,
\end{equation}
signaling a rapid extinction.

Interestingly, one can express the extinction action (\ref{deltaSsimple})
in terms of an \textit{effective} universal Hamiltonian
that is catastrophe-free. Indeed, Eq.~(\ref{deltaSsimple}) can be
interpreted as the area of a right-angled triangle:
\begin{equation}\label{rewrite}
    {\cal S}(T) \simeq \frac{B-1}{2}\, n_{eff} (T)\,,
\end{equation}
where $(B-1)$, the absolute value of $p$ at the fluctuation point
$F$ (for $B-1\ll 1$), is one of the legs of the triangle. The other
leg, $n_{eff}(T)$, is the harmonic mean of $n_s$ and $n(T)$:
\begin{equation}
\nonumber
\frac{2}{n_{eff}(T)} = \frac{1}{n_s} +\frac{1}{n(T)}\,.
\end{equation}
[To recall, close to the bifurcation $n_s\simeq N(B-1)$ is the
steady-state pre- and post-catastrophe population size and
$n(T)\simeq n_s \exp(- T)< n_s$ is the population size immediately
after the catastrophe, as predicted by the deterministic rate
equation.]  For short catastrophes $n(T)$ is close to $n_s$, and one
obtains a small correction to the unperturbed action. For long
catastrophes $n(T) \ll n_s$, and $n_{eff}(T) \simeq 2 n(T)$. We are
unaware of a simpler way to arrive at these results via
arguments based on the deterministic description of the catastrophe,
Eq.~(\ref{mfcatast}), and on the knowledge of the unperturbed
action, Eq.~(\ref{area}).

For the eikonal approximation to be
valid, we have to demand that both the total action, and the
correction to it caused by the catastrophe, be much larger than
unity. This yields a range of rescaled catastrophe durations
\begin{equation}\label{validcrit}
{\cal S}_0^{-1} \ll T\ll \ln {\cal S}_0
\end{equation}
which becomes broader as ${\cal S}_0\gg 1$ increases.

The eikonal theory provides, with exponential accuracy \cite{math}, the
value of the increase of
the extinction probability $\Delta{\cal P}_0\equiv {\cal P}_0^*-{\cal P}_0^{B}$  (see Section \ref{intro}) in terms of the ``catastrophe action" $S(T)$:
\begin{equation}\label{P0}
\Delta{\cal P}_0 \propto e^{-{\cal S}(T)}\,.
\end{equation}

\section{Comparison to numerical solutions of the master equation}
\label{numerics}

\begin{figure}
\includegraphics[width=6.4cm,height=5.2cm,clip=]{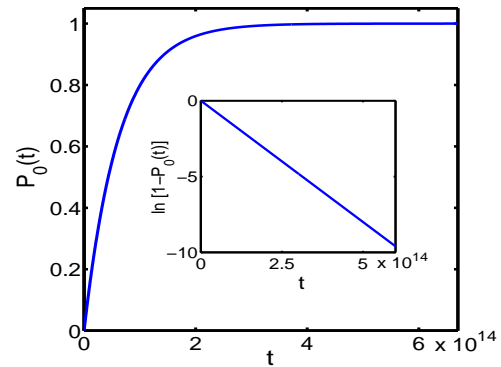}
\caption{(Color online) The extinction probability ${\cal P}_0(t)$
vs. time if there is no catastrophe. The parameters are $N=10,800$
and $B=1.08$. Inset: $\ln[1-{\cal P}_0(t)]$ vs. time. The slope of
this graph yields the MTE: $\tau\simeq 6.2 \times 10^{13}$, so that
$\ln \tau\simeq 31.8$. For comparison, the eikonal theory, see
Eq.~(\ref{SV}), predicts $\ln \tau \simeq 30.4$.}
\label{probdistext}
\end{figure}

We tested the predictions of the eikonal theory by solving the
(truncated) master equation (\ref{master}) numerically. We observed that, if there is no catastrophe, the system approaches a
quasi-stationary probability distribution after a time ${\cal
O}(\tau_0)$, as expected. This probability distribution then very slowly decays
in time, while the extinction probability ${\cal P}_0(t)$ very
slowly grows. The corresponding decay/growth time is in good
agreement with the MTE $\tau$ predicted from Eq.~(\ref{SV}), see
Fig. \ref{probdistext}.

We modeled the catastrophe by multiplying the unperturbed
reproduction rate by $f(t)$, with $f(t)$ from either
Eq.~(\ref{gencat}), or Eq.~(\ref{f}). The validity of the eikonal approximation
demands ${\cal S}_0 \gg 1$. [One must
also demand that the expected population size in the
quasi-stationary state be much larger than unity:
$q_{mf}=N(B-1)/B\gg 1$. However, above the bifurcation point $B>1$,
when the deterministic version of the Verhulst model has a
non-trivial attracting fixed point, the criterion ${\cal S}_0\gg 1$
is always more restrictive.] One more criterion for the validity of
the eikonal approximation is ${\cal S}(T)\gg 1$, for all values of
$T$ that we used.

\begin{figure}
\includegraphics[width=6.5cm,height=6.5cm,clip=]{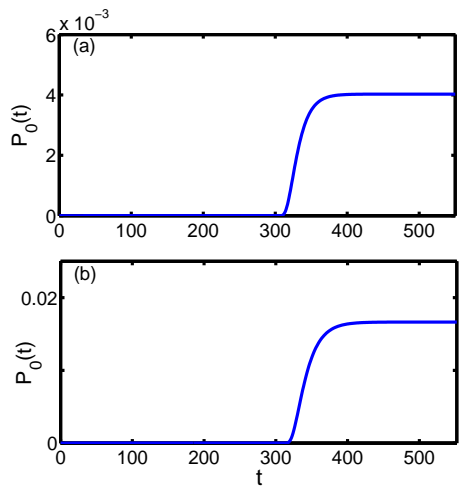}
\caption{(Color online) The extinction probability ${\cal P}_0(t)$
vs. time (at times much smaller than the MTE $\tau$), found by
solving numerically the (truncated) master equation (\ref{master1})
with $f(t)$ from Eq.~(\ref{f}). (a): $N=14,400$, $B=1.08$ and
$T=2.5$. (b): $N=10,800$, the rest of parameters is the same as in
(a). The starting time of the catastrophe in both cases, $t_c=300$,
obeys the condition $t_r\ll t_c \ll \tau$. ${\cal P}_0(t)$ before
the catastrophe  is negligibly small and cannot be seen in this
scale.} \label{p0}
\end{figure}

In our numerical solutions of the master equation the initial
condition corresponded to a fixed population size: ${\cal
P}_n(t=0)=\delta_{n,n_0}$, where $\delta_{n,n_0}$ is the Kronecker
delta, and $n_0\gg 1$. In this case an immediate extinction before
relaxing to the quasi-stationary state has an exponentially small
probability. The catastrophe time $t_c$ was chosen to be several
times longer than the relaxation time $\tau_0$ for the chosen
parameters, but \textit{much} shorter than the expected MTE.
Finally, the catastrophe duration $T$ was chosen such that the
correction to the action caused by the catastrophe and the total
action satisfy the conditions ${\cal S}_0-{\cal S}(T)\gg 1$ and
${\cal S}(T)\gg 1$. In the exactly soluble case, presented at the
end of the previous section, these two inequalities reduce to the
double inequality (\ref{validcrit}). Figure \ref{p0} presents,  for
two different sets of parameters, our numerical results for ${\cal
P}_0(t)$. Here $f(t)$ is given by Eq.~(\ref{f}).

\begin{figure}
\includegraphics[width=7.0cm,height=5.3cm,clip=]{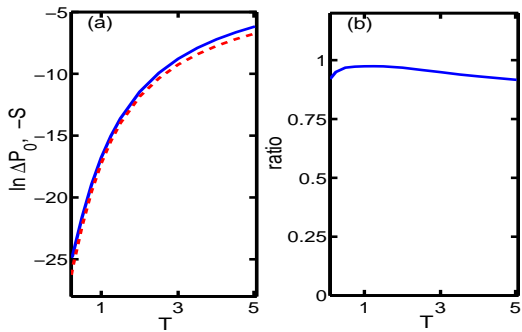}
\caption{(Color online) (a) A comparison between the natural
logarithm of the net contribution $\Delta {\cal P}_0$ of the
catastrophe to the extinction probability ${\cal P}_0(t)$ (see
text), and $-{\cal S}(T)$ from Eq.~(\ref{totact}) vs. $T$. (b) The
ratio of the two quantities vs. $T$. The parameters are $N=200$ and
$B=2$, $\Delta B=0.75$, and $t_c-t_{in}=40$.} \label{new}
\end{figure}

\begin{figure}
\includegraphics[width=7.5cm,height=6.4cm,clip=]{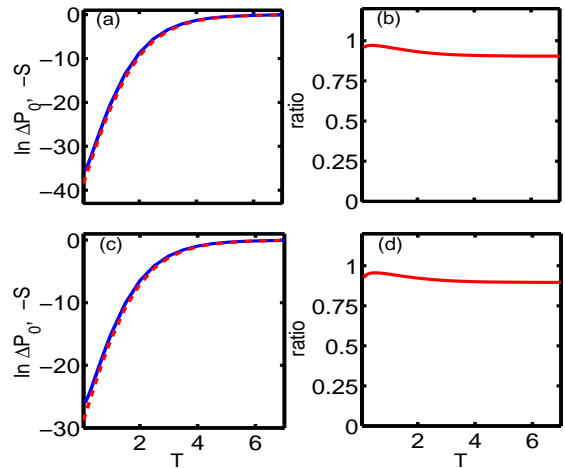}
\caption{(Color online) The natural logarithm of the numerically
computed $\Delta{\cal P}_0$ vs. $T$ (solid line) is compared to
$-{\cal S}(T)$ predicted by Eq.~(\ref{deltaSsimple}) (dashed line).
The parameters in (a) and (c) are the same as in Fig. \ref{p0} (a)
and (b), respectively. In (b) and (d) the \textit{ratios} of the
same quantities, $\ln \Delta{\cal P}_0$ and $-{\cal S}(T)$, are
plotted vs. $T$ for the same parameters as in (a) and (c),
respectively.  The analytical result in (b) deviates from the
numerical result by $2.8\%$ at the maximum. The corresponding
deviation in (d) is $4.4\%$ indicating that, as ${\cal S}_0$
increases, the agreement between the analytical and numerical
results improves.} \label{pstar}
\end{figure}

Figures \ref{new} and \ref{pstar} compare the predictions of our
eikonal theory with the numerical solutions of the master equation.
To estimate the direct impact of the catastrophe on the extinction
probability we calculated the quantity $\Delta{\cal P}_0\equiv {\cal
P}_0^*-{\cal P}_0^{B}$, where ${\cal P}_0^{B}$ is the measured
extinction probability before the catastrophe starts but after the
relaxation stage ends. As $t_c \ll \tau$, most of
the contribution to ${\cal P}_0^{B}$ comes, for the parameters and
initial conditions we worked with, from the projections of the
initial condition on the (rapidly decaying) higher eigenmodes of the
system, see \cite{Assaf2,Assaf3}.

Our numerical results for $f(t)$ from Eq.~(\ref{gencat}) are
presented in Fig. \ref{new}. The figure compares $\ln\Delta{\cal
P}_0(T)$ and $-{\cal S}(T)$ found by solving the eikonal equations
numerically, see Section $3.1$. Figure \ref{pstar} corresponds to
the simple catastrophe described by Eq.~(\ref{f}). Here $\ln
\Delta{\cal P}_0 (T)$ is compared to $-{\cal S}(T)$ for the two sets
of parameters used in Fig. \ref{p0}. These two sets of parameters
were chosen to obey the strong inequality $B-1\ll 1$, so that we
could test the analytical prediction (\ref{deltaSsimple}). Good
agreement between the theory and numerical computations is observed
in all cases, over a broad range of $T$. One can see from Fig.
\ref{new} (b) and Figs. \ref{pstar} (b) and (d) that the agreement
is best for intermediate values of $T$ and deteriorates
for small and large values of $T$. This behavior is consistent with
the validity criteria of the eikonal approximation: for too a small
$T$ the inequality ${\cal S}_0-{\cal S}(T)\gg 1$ does not hold,
whereas for too a large $T$ the inequality ${\cal S}(T)\gg 1$ does not
hold. The quantity $-(\ln \Delta{\cal P}_0)/{\cal S}$ includes
information about the pre-exponent $A(T)$
of the complete relation $\Delta{\cal P}_0 =A(T) e^{-{\cal S}(T)}$.
This pre-exponent has not yet been determined theoretically.

\begin{figure}
\includegraphics[width=7.1cm,height=6.5cm,clip=]{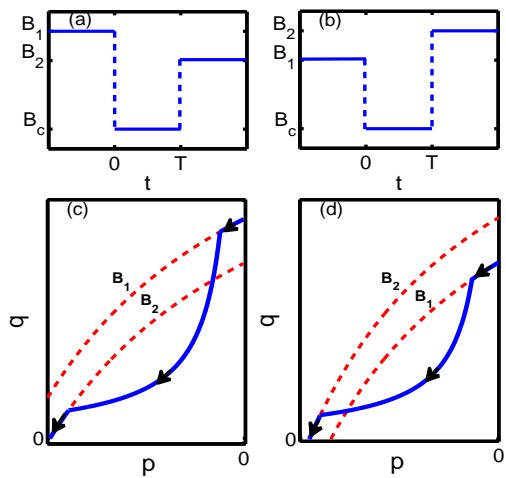}
\caption{(Color online) Not fully reversible catastrophes, as
reflected in the reproduction  rate vs. time, and the extinction
instantons for $B_1>B_2$ (a and c) and $B_1<B_2$  (b and d).  The
solid lines in (c) and (d) are the instantons, the dashed lines are
the zero-energy lines, see Eq.~(\ref{q(p)}), for $B=B_1$ and
$B=B_2$.} \label{othercat}
\end{figure}

\section{Summary and Discussion}\label{discussion}
We have employed a model example of the stochastic Verhulst logistic model to show that a catastrophic event (modeled as a temporary drop
of the reproduction rate) exponentially
increases the extinction probability of an isolated self-regulated
stochastic population. This was achieved by combining the probability generating
function technique with an eikonal approximation, and by calculating
the eikonal action over the instanton of the effective
(time-dependent) classical Hamiltonian of the system.
We have shown that eikonal approximation to the evolution equation
for the probability generating function provides a robust and
efficient way of evaluating the impact of catastrophic changes,
reflected in the reproduction/mortality rates, on the stochastic population
dynamics. Once the corresponding large parameter is present in the problem,
the eikonal approximation can be applied  to a broad class of stochastic population model, single- or multiple-species.

Although we dealt with fully
reversible catastrophes, the eikonal method is easily extendable to more
complicated situations. For instance, consider the case when the
reproduction rate $B$ reaches, after a catastrophe ends, a value
$B_2$ different from the pre-catastrophe value $B_1$. The extinction
instantons, and the corresponding total actions, can be easily found
for $B_1>B_2$ and for $B_1<B_2$, see Fig. \ref{othercat}. This
immediately yields, with exponential accuracy, the corresponding
increase in the extinction probability caused by the catastrophe.

The theory presented in this paper is non-perturbative in
the catastrophe magnitude. For relatively weak and/or short
catastrophes one can develop an eikonal \textit{perturbation} theory
by assuming that the correction to the unperturbed action ${\cal
S}_0$ is small compared to the unperturbed action itself (but still
much larger than unity, so that the eikonal theory is valid). For a
time-periodic modulation of the reaction rates such a theory has
been recently developed in Refs. \cite{Escudero,AKM}. We checked
that, for the exactly soluble catastrophe model, presented at the
end of section \ref{catastrophe}, such a perturbation theory
correctly predicts the small-$T$ asymptote of the action,
Eq.~(\ref{deltaSshort}).

Let us put the results of this work into a more general context of
employing eikonal approximation for a description of large fluctuations in stochastic
populations, the reproduction and/or mortality rates of which depend on time,
reflecting different types of environmental variations. In the present paper we have
dealt with the case of a catastrophe, where the reproduction rate undergoes a strong
temporary drop. The papers \cite{Escudero,AKM} employed eikonal approximation to deal with rate modulations which are periodic in time
(modeling daily, monthly, or annual cycles of environmental parameters). The periodicity of the rate modulation makes it possible to apply special theoretical tools unavailable otherwise (such as the Kapitsa method, applicable when the rate modulation period is very short compared to the relaxation time towards the long-lived quasi-stationary state of the population \cite{AKM}). Finally, a variant of eikonal approximation
has been recently applied to the case where the reproduction and/or
mortality rates are modulated stochastically: by noisy environmental variations \cite{shklovskii}. The stochastic environment may be thought of as a random sequence of weak and strong catastrophic events, with a certain
probability distribution of their amplitudes and durations. This
probability distribution is sensitive to the \textit{color} of the
environmental noise. Understanding the impact of a single catastrophe, achieved
in the present work, turns out to be
vital in elucidating the role of the noise color in
the population extinction\cite{shklovskii}.

The main conclusion of this work is that eikonal approximation yields a simple, accurate and visual means of assessing the viability
of stochastic populations which experience catastrophic changes of environmental conditions.

\section*{Acknowledgments}

We thank Mark Dykman and Boris Shklovskii for useful discussions. M.~A. was
supported by the Clore Foundation; M.~A. and B.~M. were supported by
the Israel Science Foundation (Grant No. 408/08); A.~K.  was supported by the NSF grant
DMR-0405212  and by the A.~P.~Sloan foundation. M.~A. and B.~M. are
grateful to FTPI of the University of Minnesota for hospitality.


\begin{thebibliography}{99}
\bibitem{spectral} The spectral method \cite{Assaf1,Assaf2,Assaf3} also yields the pre-exponent that is
lacking in Eq.~(\ref{SV}). The pre-exponent coincides with that
found, in the same limit, in Refs. \cite{Doering,Nasell01}.
\end{thebibliography}
\end{document}